\documentclass[10pt,twocolumn,twoside] {IEEEtran}
\usepackage{cite}

\ifCLASSINFOpdf
  \usepackage[pdftex]{graphicx}
  \graphicspath{{./fig/}}
\else
  \usepackage[dvips]{graphicx}
  \graphicspath{{./fig/}}
\fi

\usepackage[cmex10]{amsmath}
\usepackage[caption=false,font=footnotesize]{subfig}
\usepackage{booktabs}
\usepackage{stfloats}

\usepackage{url}
\usepackage{booktabs}

\usepackage{algorithm}
\usepackage{algorithmicx}
\usepackage{algpseudocode}

\newcommand{\avar}{\sigma_A^2}
\renewcommand{\deg}{$^\circ$}
\renewcommand{\vec}[1]{\mathbf{#1}}
\newcommand{\mat}[1]{\mathrm{#1}}

\newcommand{\yperf}[0]{\ensuremath{1/f}}
\newcommand{\bias}[1]{b_{#1}}

\newcommand{\rrw}[1]{r_{#1}}
\newcommand{\fl}[1]{f_{#1}}
\DeclareMathOperator{\var}{var}
\DeclareMathOperator{\cov}{cov}
\renewcommand{\d}[0]{\mathrm{d}}
\newcommand{\varwn}[0]{\sigma^2_{\rm WN}}
\newcommand{\varrw}[0]{\sigma^2_{q}}
\newcommand{\varfl}[0]{\sigma^2_{w}}
\newcommand{\transp}[0]{^{T}}
\newcommand{\R}[0]{\mathbb{R}}
\newcommand{\ones}[0]{\mathbb{I}}
\newcommand{\zeros}[0]{\mathbb{O}}
\renewcommand{\deg}[0]{\ensuremath{^{\circ}}}
\newcommand{\kron}[0]{\otimes}
\newcommand{\half}[0]{\frac{1}{2}}

\newlength{\doublecolumnwidth}
\usepackage{amsfonts}

\begin{document}

\title{Effect of Carouseling\\on Angular Rate Sensor Error Processes}

\author{Jussi~Collin,~\IEEEmembership{Member,~IEEE}, Martti Kirkko-Jaakkola,~\IEEEmembership{Member,~IEEE}, and Jarmo Takala,~\IEEEmembership{Senior Member,~IEEE}
\thanks{J.~Collin and J.~Takala are with Department of Pervasive Computing, Tampere University of Technology, Finland.
E-mail: jussi.collin@tut.fi}%
\thanks{M.~Kirkko-Jaakkola is with Department of Navigation and Positioning, Finnish Geodetic Institute, Finland.}%
\thanks{\textcircled{c} 2014 IEEE. This is a post-print version of a paper published in the IEEE Transactions on Instrumentation and Measurement ( DOI 10.1109/TIM.2014.2335921 ). Personal use of this material is permitted. Permission from IEEE must be obtained for all other users, including reprinting/ republishing this material for advertising or promotional purposes, creating new collective works for resale or redistribution to servers or lists, or reuse of any copyrighted components of this work in other works.} 
}%

%
%

\markboth{Effect of Carouseling on Angular Rate Sensor Error Processes (post-print)}{~}

%

\maketitle

\begin{abstract}
\boldmath
Carouseling is an efficient method to mitigate the measurement errors of inertial sensors, particularly MEMS gyroscopes. In this article, the effect of carouseling on the most significant stochastic error processes of a MEMS gyroscope, i.e., additive bias, white noise, \yperf~noise, and rate random walk, is investigated. Variance propagation equations for these processes under averaging and carouseling are defined. Furthermore, a novel approach to generating \yperf{} noise is presented. The experimental results show that carouseling reduces the contributions of additive bias, \yperf{} noise, and rate random walk significantly compared to plain averaging, which can be utilized to improve the accuracy of dead reckoning systems.
\end{abstract}

\begin{IEEEkeywords}
\boldmath gyroscopes, microelectromechanical systems, noise generators, stochastic processes, \yperf\ noise
\end{IEEEkeywords}

%
\IEEEpeerreviewmaketitle

\section{Introduction}
%
%
%
%
\IEEEPARstart{M}{EMS} gyroscope (gyro) technology has developed rapidly during the past years, and MEMS gyros have now been shown to be able to perform even high-precision tasks such as gyrocompassing (i.e., North seeking by observing the Earth's rotation rate)~\cite{Renkoski2008Phd,Iozan1,prikhodko2013}. The small physical size, power consumption, and batch manufacturing cost make MEMS gyros ideal for a variety of applications in, e.g., land vehicles and mobile devices.

The key to high accuracies is sophisticated compensation of measurement errors which exhibit significantly more variations on MEMS gyros than in the case of, e.g., optical sensors. Common strategies are error model calibration when the true rotation is known (usually zero)~\cite{Kirkko1,Nourel1} and deliberately altering the orientation of the sensitive axis of the gyro in order to separate measurement errors from the input signal. Intentional slewing of the gyro has two main approaches: the sensor can be either rotated continuously, which is referred to as \emph{carouseling}, or it can be rotated at specific discrete intervals (often $180\deg$), which is known as \emph{indexing}~\cite{insstandard,Sandia1}. In the literature, the terms \emph{maytagging} and \emph{two-point carouseling} are also used for indexing. Turntables are commonly used for offline calibration of inertial measurement units~(IMUs), e.g.,~\cite{syed2007}, but in this article, we study IMU rotations that are applied during the actual measurement.

Rotating IMUs have been studies already since the 1960s~\cite{geller68}, and much research has been focused on high-quality sensors. However, MEMS gyros whose measurement errors are less stable than those of, e.g., ring laser gyros, can benefit even more from IMU slewing because the fluctuating error processes are significantly more difficult to estimate. Significant improvements in pedestrian dead reckoning obtained using a foot-mounted rotating IMU have been reported~\cite{hide2012rotating}, although the test setup was too cumbersome for real-life use. However, a dedicated rotating system is not always necessary, e.g., if the IMU is mounted at the wheel of a land vehicle~\cite{patentti}.

The studies of carouseling and indexing errors have been mostly investigating common inertial navigation error states such as biases and scale factors~\cite{lai2010,GaoSun,Nie1,Yuan1}. In contrast, this article studies the stochastic error components in the output of a MEMS gyro and analyzes their variance propagation under carouseling in comparison with non-carouseled averaging. In addition to MEMS sensors, the results are relevant for other types of rate sensors where non-stationary noise processes cause significant errors after integration over time.

Among the possible error processes in the output of a MEMS gyro is \yperf{} (flicker) noise whose name originates from its power spectral density. \yperf{} noise is a long-memory process and is nontrivial to synthesize~\cite{shusterman1998,kasdin1995,narasimha2005,eliazar2010,Sabatini806085,Yaz554304}. In this article, we both use the fractional integral model of \yperf{} noise~\cite{Rodriguez2009} and propose a novel approach to generating noise with constant Allan variance at certain averaging times. Synthetic \yperf{} noise can be used to simulate not only MEMS gyros but also, e.g., transient circuits~\cite{hillermeier2004} or other phenomena where \yperf{} noise is encountered~\cite{vanderziel1988,Voss1979}.

This article is organized as follows. The stochastic processes used for modeling the most important error processes in the output of MEMS gyros are defined in Section~\ref{sec:processes}, followed by carouseling analysis in Section~\ref{sec:carouseling}. A novel approach of synthesizing noise with constant Allan variance is presented in Section~\ref{sec:constavar}, and experimental validation is carried out in Section~\ref{sec:experiments}. Finally, Section~\ref{sec:conclusions} concludes the article.

\section{Error Process Models} \label{sec:processes}

In this section, the models used for different error components in the output of angular rate sensor are described. These process models are chosen based on~\cite{coriolisstandard}. Similar models have been employed, e.g., in~\cite{vaccaro2012} except for that \yperf{} noise was not considered in~\cite{vaccaro2012}. Errors that are dependent on the input signal magnitude (scale factor errors) are neglected in the following discussion.

\subsection{Additive Bias}

The additive bias~$\bias{}$ is modeled as a random constant; therefore,
\begin{equation}\label{eq:bias}
	\bias{t} = \bias{t-1}.
\end{equation}
It would be possible to embed the bias in other error processes, but in this article, other error processes are treated as zero-mean.

\subsection{White Noise}

White noise, often called \emph{angular random walk} in the context of gyroscopes, is a random process where the samples are mutually independent. It is assumed that the process has zero mean and a constant variance~$\varwn$. Many factors contribute white noise to the sensor output. For instance, quantization noise is white, and so is thermal noise.

\subsection{Rate Random Walk}

Rate random walk~(RRW) is the sum of independent and identically distributed, zero-mean increments, modeled as
\begin{equation}\label{eq:rrw}
	\rrw{t} = \rrw{t-1} + q_t
\end{equation}
where $q_1$, $q_2$, \ldots\ constitute a white noise process with variance~$\varrw$. It can be seen that RRW is a Markov process, i.e., memoryless: the value~$\rrw{t}$ does not depend on other previous or future realizations of the process than~$\rrw{t-1}$. An important source of RRW are changes in the temperature of the sensor. Other than that, RRW is caused by, e.g., aging of the sensor element.

\subsection{\yperf\ Noise}

In this article, \yperf\ noise is modeled as a fractional integral of a white noise sequence. Assuming a degree of integration~$0 < d < 1$, \yperf\ noise is modeled in discrete time as~\cite{hosking81}
\begin{equation}\label{eq:yperf}
	\fl{t} = \sum_{i=1}^t \frac{\Gamma\left(t-i+d\right)}{\Gamma\left(t-i+1\right)\Gamma(d)} w_i
\end{equation}
where $w_1$, $w_2$, \ldots\ are white noise with variance~$\varfl$ and $\Gamma$~denotes the gamma function. As opposed to RRW, \yperf{} noise is a long-memory process~\cite{keshner82}. Although encountered in various contexts, its physical origin is unknown~\cite{Voss1979}.

\section{Gyroscope Carouseling}\label{sec:carouseling}

\begin{figure}
	\centering
	\includegraphics[scale=1]{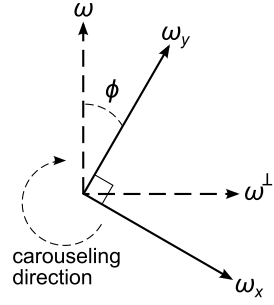}
	\caption{Schematic of gyroscope carouseling}
	\label{fig:carouseling_schematic}
\end{figure}

In this article, the concept of carouseling is defined as follows; a schematic is shown in Fig.~\ref{fig:carouseling_schematic}. Consider two gyros with sensitive axes~$x$ and~$y$ aligned perpendicular to each other. In carouseling, these sensors are intentionally rotated on the plane defined by the sensitive axes, and the measurements~$\omega_x$ and~$\omega_y$---deteriorated by biases, noise, and other imperfections---are used to estimate the true angular rate~$\omega$ about a fixed ``virtual'' axis~$\phi = 0$ on the plane of the sensitive axes. The outputs of the gyros are combinations of the angular rate of interest~$\omega$ and the angular rate about the axis~$\phi = 90\deg$, denoted here by~$\omega^\perp$:
\begin{subequations}\label{eq:gyrocomponents}
	\begin{align}
	\omega_x & =  -\omega\sin\phi + \omega^\perp\cos\phi + \epsilon_x \\
	\omega_y & =  \phantom{-}\omega\cos\phi + \omega^\perp\sin\phi  + \epsilon_y
	\end{align}
\end{subequations}
where $\epsilon_x$ and $\epsilon_y$ denote additive sensor measurement errors as a superposition of the error processes defined in Section~\ref{sec:processes}. The instantaneous angular rate at the orientation of interest is then estimated as
\begin{equation}\label{eq:instantaneouscarousel}
	\widehat{\omega} = -\omega_x \sin\phi + \omega_y \cos\phi = \omega + \widetilde{\epsilon};
\end{equation}
the angular rate~$\widehat{\omega}^\perp$ about the axis perpendicular to the axis of interest could be computed in a similar manner, but, in this article, we focus on the analysis of a single axis of interest.

Since the sensors can only be sampled at discrete intervals, only discrete values of the carouseling angle~$\phi$ need to be considered. In the analysis presented in this article, we assume a uniform carouseling rate with period~$T$~seconds and a uniform sensor sampling frequency~$N/T$~hertz for an integer~$N$. We will focus on the average angular rate during the $t$th carouseling revolution which is estimated as
\begin{equation}\label{eq:carousel}
\begin{split}
	\omega_t = \frac{1}{N} \sum_{i=1}^{N}	 &-\omega_x\left((t-1)T + i\frac{T}{N}\right) \sin \frac{2\pi i}{N} \\
	&+ \omega_y\left((t-1)T + i\frac{T}{N}\right) \cos \frac{2\pi i}{N} .
\end{split}
\end{equation}
In this section, the effect of carouseling on the measurement errors present in~$\omega_x$ and~$\omega_y$ is studied in terms of the stochastic processes defined in Section~\ref{sec:processes}. Comparison is made to a single gyroscope that is not carouseled, i.e., whose measurements are directly averaged over intervals of $T$~seconds.

\subsection{Additive Bias} \label{sec:biasanalysis}

For simplicity, consider only the behavior of one of the two (physical) gyros during a carouseling revolution. In continuous time, it is obvious that carouseling cancels the bias because
\begin{equation}\label{eq:continuoustimebias}
	\int_0^{2\pi} \bias{}\sin \phi\ \d\phi = 0
\end{equation}
and the same applies to the other gyro with cosine coefficients. Fortunately, this is the case in discrete time as well under the assumption that the sampling rate is uniform and an integer multiple of the carouseling frequency. Consider the sum
\begin{equation}
	\frac{1}{N}\sum_{i=1}^{N} \bias{} \sin \frac{2\pi i}{N} = \frac{\bias{}}{N} \sum_{i=1}^{N} \sin \frac{2\pi i}{N} .	
\end{equation}
Using Euler's formula, the sum can be interpreted as the imaginary part (or the real part in the case of cosine terms) of the sum of $N$~roots of unity which is well known to be zero for all $N > 1$. In contrast, it is clear that direct averaging has no influence on the constant additive bias.

\subsection{White Noise}

Averaging uncorrelated noise obviously decreases its variance. Computing the variance of the carouseled angular estimate yields
\begin{equation}\label{eq:wnvariance}
\begin{split}
	\var \omega_t =& \frac{1}{N^2} \sum_{i=1}^{N} \varwn \sin^2 \frac{2\pi i}{N} + \varwn \cos^2 \frac{2\pi i}{N} \\
	=& \frac{\varwn}{N}
\end{split}
\end{equation}
which is equal to the variance of the directly averaged white noise sequence. Therefore, carouseling does not have advantages over direct averaging in terms of white noise.

\subsection{Rate Random Walk}\label{sec:rrwanalysis}

In order to analyze the joint distribution of two consecutive carouseling revolutions, define the $N\times N$ lower triangular cumulative sum matrix
\begin{equation}
	\mat{R} = \begin{bmatrix}
		1 & 0 & \cdots \\
		1 & 1 & \ddots \\
		\vdots & \vdots & \ddots
	\end{bmatrix},
\end{equation}
and partition the corresponding $2N\times 2N$ cumulative sum matrix as
\begin{equation}
	\mat{R_2} = \begin{bmatrix}
		\mat{R} & \zeros  \\
		\ones & \mat{R}
	\end{bmatrix}
\end{equation}
where $\zeros$ and $\ones$ denote $N\times N$ matrices of zeros and ones, respectively.
Also define the integrator vector
\begin{equation}
	\vec{1} = \frac{1}{N}\begin{bmatrix}1 & 1 & \cdots\end{bmatrix}\transp \in \R^N ,
\end{equation}
and the carouseling coefficient vectors
\begin{equation}
	\begin{split}
	\vec{s} &= \frac{1}{N}\begin{bmatrix} \sin 2\pi\frac{1}{N} & \sin 2\pi\frac{2}{N} & \cdots & \sin 2\pi\end{bmatrix}\transp \\
	\vec{c} &= \frac{1}{N}\begin{bmatrix} \cos 2\pi\frac{1}{N} & \cos 2\pi\frac{2}{N} & \cdots & \cos 2\pi\end{bmatrix}\transp .
\end{split}
\end{equation}

Now, given a white noise vector~$\vec{q}\in\R^{2N}$, two consecutive direct $N$-averages of a RRW sequence would be obtained as
\begin{equation}\label{eq:consecutiveaverages}
	\begin{bmatrix}
	\vec{1}\transp & \vec{0}\transp \\
	\vec{0}\transp & \vec{1}\transp
	\end{bmatrix} \mat{R_2} \vec{q}
\end{equation}
where $\vec{0}$ is a $N\times 1$ vector of zeros. As the covariance of~$\vec{q}$ is, by definition, $\varrw\mat{I}$ where $\mat{I}$ denotes the identity matrix, the covariance matrix of~\eqref{eq:consecutiveaverages} is computed as
\begin{equation}\label{eq:averagecoefficient}
	\begin{split}
	&\begin{bmatrix}
	\vec{1}\transp & \vec{0}\transp \\
	\vec{0}\transp & \vec{1}\transp
	\end{bmatrix} \mat{R_2}\, \varrw\mat{I}\, \mat{R_2}\transp
	\begin{bmatrix}
	\vec{1} & \vec{0} \\
	\vec{0} & \vec{1}
	\end{bmatrix} \\
	&=  \varrw\begin{bmatrix} \vec{1}\transp \mat{R}\mat{R}\transp\vec{1} & \vec{1}\transp \mat{R}\ones\transp\vec{1} \\
	 \vec{1}\transp \ones\mat{R}\transp\vec{1} & \vec{1}\transp \mat{R}\mat{R}\transp\vec{1} + \vec{1}\transp \ones\ones\transp\vec{1}
	 \end{bmatrix} \\
	&=  \varrw \begin{bmatrix} \frac{1}{N^2}\sum_{i=1}^N i^2 & \frac{1}{N^2}\sum_{i=1}^N iN \\
	\frac{1}{N^2}\sum_{i=1}^N iN & \frac{1}{N^2}\sum_{i=1}^N i^2 + \frac{N^3}{N^2}
	\end{bmatrix} \\
	&=  \varrw\begin{bmatrix} \frac{2N^3 + 3N^2 + N}{6N^2} & \frac{N+1}{2} \\
	\frac{N+1}{2} & \frac{2N^3 + 3N^2 + N}{6N^2} + N
	\end{bmatrix} .
	 \end{split}
\end{equation}
It can be seen that the two averages are correlated and that the variance of the second average is approximately proportional to $4N/3$. Every subsequent average will have variance higher by~$\varrw N$ than the previous one, which is intuitively understood because of the process model~\eqref{eq:rrw} and can be seen by repeating the calculations for~$\mat{R}_3$ etc. A rigorous proof by induction is, however, not given here.

Analogously, the carouseled averages would be computed as
\begin{equation}\label{eq:consecutivecarousels}
	-\begin{bmatrix}
	\vec{s}\transp & \vec{0}\transp \\
	\vec{0}\transp & \vec{s}\transp
	\end{bmatrix} \mat{R_2} \vec{q}_x +
	\begin{bmatrix}
	\vec{c}\transp & \vec{0}\transp \\
	\vec{0}\transp & \vec{c}\transp
	\end{bmatrix} \mat{R_2} \vec{q}_y
\end{equation}
where both gyros have their respective realizations of RRW driving noise. We will assume that the RRW increments~$q_t$ of the two gyros are statistically independent and identically distributed~(i.i.d.). In principle, the increments are correlated at least for the component caused by changes in the ambient temperature, but as long as the carouseling period~$T$ is reasonably short (e.g., in the order of $1$~second), temperature fluctuations during one carouseling revolution can be neglected in most applications. If the sensors are of identical make and model and originate from the same production batch, the assumption of identical distributions can be considered reasonable.

Keeping in mind that the elementwise sums of the vectors~$\vec{s}$ and~$\vec{c}$ equal~$0$ as discussed in Section~\ref{sec:biasanalysis}, i.e., $\vec{1}\transp \vec{s} = \vec{1}\transp \vec{c} = 0$, the covariance of the sine coefficient term in~\eqref{eq:consecutivecarousels} is expressed as
\begin{equation}\label{eq:carouselcoefficient}
\begin{split}
&\varrw\begin{bmatrix} \vec{s}\transp \mat{R}\mat{R}\transp\vec{s} & \vec{s}\transp \mat{R}\ones\transp\vec{s} \\
	 \vec{s}\transp \ones\mat{R}\transp\vec{s} & \vec{s}\transp \mat{R}\mat{R}\transp\vec{s} + \vec{s}\transp \ones\ones\transp\vec{s}
	 \end{bmatrix} \\
	 &= \varrw\begin{bmatrix} \vec{s}\transp \mat{R}\mat{R}\transp\vec{s} & 0 \\
	 0 & \vec{s}\transp \mat{R}\mat{R}\transp\vec{s}
	 \end{bmatrix}
	\end{split}
\end{equation}
which implies that the consecutive carouseled averages are uncorrelated and have equal variances. To compute the values of these variances, let us interpret $\vec{s}\transp \mat{R}\mat{R}\transp\vec{s}$ as a numerical integration according to the rectangle rule:
\begin{equation}\label{eq:carouselcoefficientvalue}
	\begin{split}
	\vec{s}\transp \mat{R}\mat{R}\transp\vec{s} =& N \frac{1}{N}\sum_{i=1}^N\left(\frac{1}{N}\sum_{j=i}^N \sin\frac{2\pi k}{N}\right)^2 \\
	\approx& N \int_0^1\left(\int_x^1 \sin 2\pi y\, \d y\right)^2 \d x \\
	&= \frac{N}{2\pi} \int_0^1\left(\cos 2\pi x - 1\right)^2\d x = \frac{3N}{8\pi^2} .
	\end{split}
\end{equation}
Similar computations for the cosine term yield an asymptotic proportionality coefficient of $N/(8\pi^2)$, summing up to a total variance of $\varrw N/(2\pi^2)$. This is already $96~\%$ smaller than the asymptotic coefficient~$4/3$ obtained for direct averaging in~\eqref{eq:averagecoefficient}, and by repeating the calculations for $\mat{R}_3, \ldots$ one can see that subsequent carouseled averages have the same variance as opposed to direct averaging where the variances increase linearly. A rigorous proof is again omitted, but the phenomenon can be intuitively understood based on the Markov property of RRW and the result obtained in Section~\ref{sec:biasanalysis}.

Note that the carouseling period~$T$ does not appear explicitly in~\eqref{eq:averagecoefficient} and~\eqref{eq:carouselcoefficientvalue}. However, as long as the sensor sampling rate is constant, the carouseling period~$T$ is proportional to the number of carouseling points~$N$, and on the other hand, the variance~$\varrw$ of the RRW driving noise depends on the sampling rate.

\subsection{\yperf\ Noise}

Analogously to the analysis presented for RRW in Section~\ref{sec:rrwanalysis}, define the matrix~$\mat{F} \in \R^{N\times N}$ which produces a \yperf{} noise sequence by multiplying a white noise sequence~$\vec{w}$. This matrix is unit lower triangular with subdiagonal entries computed according to~\eqref{eq:yperf}; in fact, it is easy to see that $\mat{F}$ is also a Toeplitz matrix. Consequently, the product~$\mat{F}\vec{1}$ effectively computes the cumulative sum of the first column of~$\mat{F}$, and $\mat{F}\transp\vec{1}$ contains the same values in the reverse order. To analyze the convergence of this sum, let us conduct a limit comparison test and compute the limit of the ratio of $i^{d-1}$ and the $i$th entry of the first row of $\mat{F}\transp$ as $i$ tends to infinity:
\begin{equation}\label{eq:limitcomparisontest}
\begin{split}
	\lim_{i \to \infty} \frac{\Gamma\left(i+d\right)}{\Gamma\left(i+1\right)\Gamma(d)} \Big/ i^{d-1}
	&=
\lim_{i \to \infty} \frac{\Gamma\left(i+d\right) i^{1-d}}{i \Gamma(i)\Gamma(d)} \\
 &=\lim_{i \to \infty} \frac{\Gamma\left(i+d\right)}{i^{d} \Gamma(i)\Gamma(d)} \\
 &= \frac{1}{\Gamma(d)} > 0 \quad \forall \ d > 0.
\end{split}
\end{equation}
Since the series~$\sum_{i=0}^\infty i^{d-1}$ is well known to diverge for all~$d \geq 0$, the limit comparison test concludes that the elements in the product~$\mat{F}\vec{1}$ also tend to infinity with increasing time~$t$ and positive~$d$. 

\begin{figure}[t]
	\centering
	\includegraphics[width=\doublecolumnwidth]{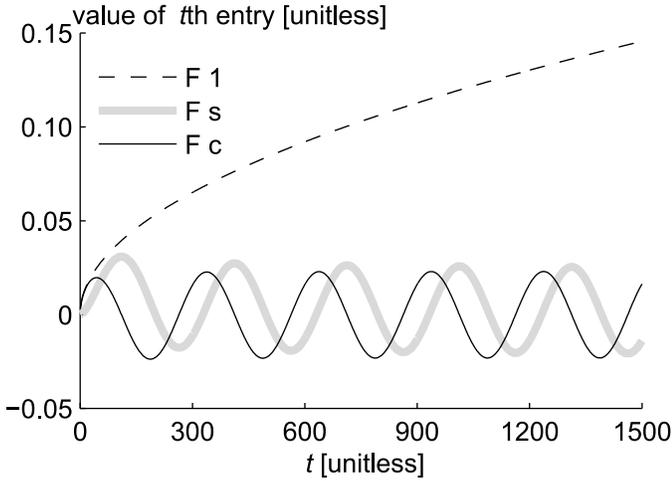}
	\caption{Effect of carouseling on \yperf~noise}
	\label{fig:yperf_cumsum}
\end{figure}

As opposed to the strictly positive direct averaging vector~$\vec{1}$, the carouseling coefficient vectors~$\vec{s}$ and~$\vec{c}$ contain both positive and negative entries. In fact, if $N$ happens to be even, these vectors contain $N/2$ pairs of opposite numbers. Then, the cumulative carouseled sum $\mat{F}\transp\vec{s}$ or $\mat{F}\transp\vec{c}$ can be interpreted, as the dimension of the matrix~$\mat{F}$ increases, as the sum of $N/2$ different alternating series. The absolute values of the terms in these series are decreasing and, therefore, these alternating series converge according to the Leibniz criterion. Fig.~\ref{fig:yperf_cumsum} illustrates the behavior of carouseled and directly averaged \yperf~noise with $d = 1/2$ and $N = 300$. As the values in $\mat{F}\transp\vec{1}$ grow larger than those in $\mat{F}\transp\vec{s}$ and $\mat{F}\transp\vec{c}$, it can be expected that the variances computed using the squared form $\varfl\vec{1}^T\mat{F}\mat{F}\transp\vec{1}$ also increase significantly faster than their carouseled counterparts.  

\section{Constant Allan Variance} \label{sec:constavar}

A popular tool for analyzing gyroscope measurement errors is the Allan variance. It is especially suitable for the analysis of indexing because both of them operate based on the differences of consecutive sample averages. The Allan variance~$\avar(\tau)$ is a function of averaging time~$\tau$, computed as
\begin{equation} \label{eq:avar}
	\avar( \tau ) = \frac{1}{2\left( M-1\right)}\sum_{j=1}^{M-1}
\left(\bar{y}(\tau)_{j+1} - \bar{y}(\tau)_{j} \right)^2
\end{equation}
where the values of $\bar{y}(\tau)_{j}$ and $M$ are obtained by dividing the data~$y$
into disjoint bins of length~$\tau$, $\bar{y}(\tau)_{j}$ is the average
value of the $j$th bin, and $M$ is the total number of
bins~\cite{Kirkko1,AllanSTD}. When defined this way, $\avar( \tau )$ is  a
statistic, function of a gyro noise sample~$y$. If $y$ consists of pure \yperf{}
noise, it can be expected that $\avar( \tau )$ is independent of
$\tau$~\cite{rlgstandard,Voss1979,Greenhall97}. In this section, we introduce a discrete
sequence that has this property; the term `sequence' is used instead of `stochastic process' because the data are generated in a non-causal procedure and the variance of the individual random variables in the sequence is a function of the length of the sequence.

\begin{algorithm}
	\caption{Generating the deterministic sequence~$S_{2^n}$}
	\label{alg:deterministic}
	\begin{algorithmic}[1]
			\Require $1 < n \in \mathbb{N}$
			\Ensure $\vec{v} = S_{2^n}$
			\State $\vec{v} = \left[-\frac{1}{2} \quad \frac{1}{2}\right]$
			\For{$i = 2, \ldots, n$}
				\State $\vec{v} = \vec{v} \kron \left[1\quad 1\right] + \vec{a}_{1\ldots 2^i}$ \label{alg:deterministic:hline}
			\EndFor
	\end{algorithmic}
\end{algorithm}

\begin{figure}
	\centering
	\includegraphics[width=\doublecolumnwidth]{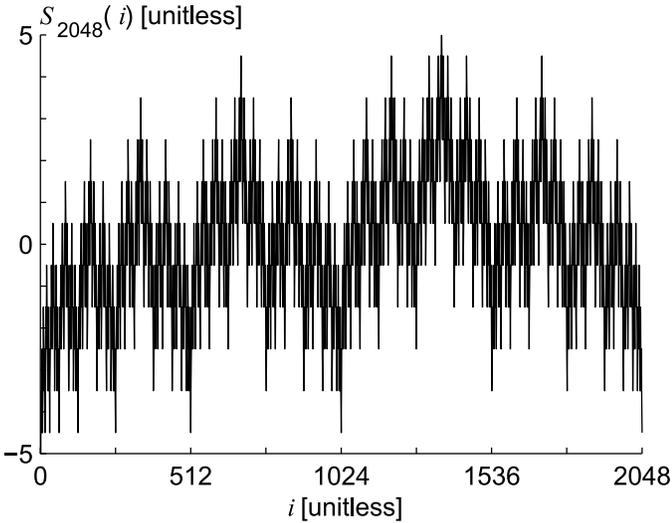}
	\caption{The sequence~$S_{2048}$ as generated using Algorithm~\ref{alg:deterministic}}
	\label{fig:s2048}
\end{figure}

Algorithm~\ref{alg:deterministic} describes a procedure to generate a sequence~$S_{2^n}$ of length~$2^n$ for which
\begin{equation} \label{eq:avars1}
\left| \bar{S}_{2^n}(\tau)_{j+1} - \bar{S}_{2^n}(\tau)_{j} \right|=1, \forall
\tau=1,2,4,\ldots, 2^{n-1};
\end{equation}
the progress of the algorithm is tabulated in Table~\ref{tblS}.
Starting with the sequence~$S_2 = \left[ -\frac{1}{2},\frac{1}{2}\right]$, use the Kronecker product~$\kron$ to duplicate the elements of~$S_2$ to
obtain $\left[-\frac{1}{2},-\frac{1}{2},\frac{1}{2},\frac{1}{2}\right]$. Clearly, \eqref{eq:avars1} now holds for $\tau = 2$; in order to make it valid for $\tau = 1$ as well, add the sequence~$\vec{a}_{1\ldots 2^i}$ to the Kronecker product where
\begin{equation}\label{eq:signs}
a_k = \begin{cases}
			 -\frac{1}{2} & \text{if $k = 1$} \\
			 +\frac{1}{2} & \text{if $k = 2$} \\
			 -a_{k-2} & \text{otherwise.}
			 \end{cases}
\end{equation}
It is easy to see that the elements of~$\vec{a}_{1\ldots 2^i}$ repeat with a period of four. The resulting sequence is $S_4 = \left[ -1,0,1,0\right]$. As an example, the sequence~$S_{2048}$ is plotted in Fig.~\ref{fig:s2048}. The sequence is quantized at distinct values because it is generated as a superposition of~$\log_2 2048 = 11$ square waves. According to~\cite{keshner82}, a logarithmic amount of state variables is sufficient to characterize a \yperf{} process.

\begin{table}
\centering
\caption{Generating the deterministic sequence~$S_8$}
\label{tblS}
\begin{tabular}{ l c c c c c c c c}
\toprule
$S_2$&\multicolumn{4}{c}{$-\half$}   & \multicolumn{4}{c}{$\half$}   \\
\midrule[\heavyrulewidth]
$\kron [1~1]$&\multicolumn{2}{c}{$-\half$}   & \multicolumn{2}{c}{$-\half$} &
\multicolumn{2}{c}{\phantom{$+$}$\half$} &   \multicolumn{2}{c}{\phantom{$-$}$\half$}\\
$\vec{a}_{1\ldots 4}$&\multicolumn{2}{c}{$-\half$}   & \multicolumn{2}{c}{$+\half$} &
\multicolumn{2}{c}{$+\half$} &   \multicolumn{2}{c}{$-\half$}\\
\midrule
$S_4$   &\multicolumn{2}{c}{$-1$}   & \multicolumn{2}{c}{\phantom{$+$}$0$}
& \multicolumn{2}{c}{\phantom{$+$}$1$} &   \multicolumn{2}{c}{\phantom{$+$}$0$}\\
\midrule[\heavyrulewidth]
$\kron [1~1]$   &  $-1$ &  $-1$   & \phantom{$+$}$0$ & \phantom{$-$}$0$    &  \phantom{$-$}$1$  & \phantom{$+$}$1$ &
\phantom{$-$}$0$ &\phantom{$-$}$0$  \\
$\vec{a}_{1\ldots 8}$   &  $-\half$ &  $+\half$   & $+\half$  &$-\half$    &  $-\half$  & $+\half$ & $+\half$ &
$-\half$  \\
\midrule
   $S_8$& $-\frac{3}{2}$     & $-\half$  &\phantom{$+$}$\half$    &   $-\half$
&\phantom{$+$}$\half$ & \phantom{$+$}$\frac{3}{2}$ & \phantom{$+$}$\half$ & $-\half$ \\
\bottomrule
\end{tabular}
\end{table}

Similarly, to obtain a stochastic sequence~$R$, replace the deterministic
value~$1$ in~\eqref{eq:avars1} by the random variable $x_\tau$ with zero mean and variance~$C^2$:
\begin{equation} \label{eq:avars1stoc}
\bar{R}(\tau)_{j+1} - \bar{R}(\tau)_{j}=x_\tau, \forall
\tau=1,2,4,\ldots, 2^{n-1}. \end{equation}
To obtain such a sequence, draw~$x_i$
with zero mean and unit variance and add $\vec{a}_{1\ldots 2^i} x_i$
instead of $\vec{a}_{1\ldots 2^i}$ to the Kronecker product at line~\ref{alg:deterministic:hline} of Algorithm~\ref{alg:deterministic}. This procedure is tabulated in Table~\ref{tab:stocseq}.
\begin{table}[t]
\centering
\caption{Generating the stochastic sequence~$R_4$}
\label{tab:stocseq}
\begin{tabular}{ c c c c c}
\toprule
$R_2$ &\multicolumn{2}{c}{$-\half x_1$}   & \multicolumn{2}{c}{$\half x_1$}   \\
\midrule[\heavyrulewidth]
&        $-\half x_1$  &  $-\half x_1$   &   \phantom{$+$}$\half x_1$&
\phantom{$+$}$\half x_1$ \\
   &       $-\half x_2$  &  $+\half x_2$    & $+\half x_2$ & $-\half x_2$ \\
\midrule
$R_4$   &         $-\half x_1-\half x_2$    &   $-\half x_1+\half x_2$    &
$\half x_1+\half x_2$ & $\half x_1-\half x_2$ \\
\bottomrule
\end{tabular}
\end{table}
The sequence~$R_{2^n}$ can be expressed as a matrix--vector product~$\mat{K}\vec{x}$
where $\mat{K} \in \R^{2^n \times n}$ is a constant matrix and $\vec{x} \in \R^{2^n}$ is an i.i.d.\ random vector; the $i$th column of the matrix~$\mat{K}$ is computed as the Kronecker product of $\vec{a}_{1\ldots 2^i}$ and a $2^{n-i}\times 1$ vector of ones. For $n = 2$,
\begin{equation}\label{eq:Kmatrix}
\mat{K} =	\begin{bmatrix}
	-0.5 & -0.5 \\
	-0.5 & \phantom{-}0.5\\
\phantom{-}0.5 & \phantom{-}0.5\\
\phantom{-}0.5 & -0.5
	\end{bmatrix}
\end{equation}
and then
\begin{equation} \label{eq:Ktimes}
R_4=\mat{K}[x_1\quad x_2]\transp.
\end{equation}
Under the i.i.d.\ and unit variance assumptions we have
\begin{equation}\label{eq:covx}
\cov \vec{x} =	\begin{bmatrix}
	1 & 0 \\
	0 & 1\\
	\end{bmatrix}
\end{equation}
and thus
\begin{equation}\label{eq:CKmatrix}
\cov R_4=\mat{K}\mat{K}\transp=	\begin{bmatrix}
	\phantom{-}0.5 & \phantom{-}0   &           -0.5 & \phantom{-}0 \\
	\phantom{-}0   & \phantom{-}0.5 & \phantom{-}0   &           -0.5\\
              -0.5 & \phantom{-}0   & \phantom{-}0.5 & \phantom{-}0\\
	\phantom{-}0   &           -0.5 & \phantom{-}0   & \phantom{-}0.5
	\end{bmatrix}.
\end{equation}
To obtain the constant~$C^2 = \var x_\tau$ in~\eqref{eq:avars1stoc}, express the data bin average differences in~\eqref{eq:avar} in the matrix--vector product form $\mat{A}\vec{y}$ as
 \begin{equation}\label{eq:matform}
 \bar{y}(\tau)_{j+1} - \bar{y}(\tau)_j =\left[-\frac{1}{\tau} ~ \ldots ~ -\frac{1}{\tau} ~ \frac{1}{\tau} ~ \ldots ~ \frac{1}{\tau}\right] \begin{bmatrix}
 y_{(j-1)\tau + 1}\\
 \vdots\\
 y_{j\tau}\\
 y_{j\tau + 1}\\
 \vdots\\
 y_{(j+1)\tau}
 \end{bmatrix}
  \end{equation}
 to obtain
\begin{equation}\label{eq:Amatrix}
\mat{A} =	\begin{bmatrix}
-1           & \phantom{-}1 & \phantom{-}0 &0\\
\phantom{-}0 & -1           & \phantom{-}1 & 0 \\
\phantom{-}0 & \phantom{-}0 & -1           &1\\
	\end{bmatrix}
\end{equation}
and then
\begin{equation}\label{eq:AAmatrix}
\cov \mat{A}\vec{y} = \cov \mat{A}\mat{K}\vec{x} = \mat{A}\mat{K}\mat{K}\transp
\mat{A}\transp=	 \begin{bmatrix}
	\phantom{-}1 & 0 & -1  \\
	\phantom{-}0 & 1 & \phantom{-}0\\
   -1            & 0 & \phantom{-}1\\
	\end{bmatrix},
\end{equation}
showing that, indeed, the variance~$C^2 = 1$. If this variance is unknown, it
can be shown that~\eqref{eq:avar} without the term $\frac{1}{2}$ yields
an unbiased estimate of $C^2$, i.e., $E\left[2\avar( \tau )\right]=C^2$. However, it is not the
minimum-variance unbiased estimator (MVUE), which can be found by using the
Moore--Penrose pseudoinverse of~$\mat{K}$,
\begin{equation}\label{eq:Kpseudomatrix}
\mat{K}^+ = \begin{bmatrix}
	-0.5 & -0.5 & 0.5 & \phantom{-}0.5  \\
	-0.5 & \phantom{-}0.5  & 0.5 & -0.5\\
	\end{bmatrix}.
\end{equation}
Now, $\mat{K}^+R_4=[x_1~x_2]^T$, and the well known sample variance of this is the MVUE. Having a theoretical mean value for a process with constant Allan variance can help in extending the statistical models discussed in~\cite{vaccaro2012} to \yperf{}-type processes. The constant variance property was derived with nonoverlapping Allan variance and does not hold exactly for overlapping Allan variance estimators~\cite{LiOverlap,LiSliding}.

Interestingly, $\mat{K}_{16\times4}+0.5$  is equal to the standard Gray code representation in matrix
form~\cite[Table~1]{flajolet1980note}. Thus, as
  \begin{equation}
  S_m=\mat{K} \begin{bmatrix}1 & 1 & \cdots\end{bmatrix}\transp,
    \end{equation}
  the sequence $S+{n/2}$ also depicts the number of ones in the Gray code
representation, and bounds for the sequence can be obtained from number
theory~\cite{mcilroy1974number}.  Other interesting properties can be found as well:
for example, the columns of the covariance matrix of the sequence~$\left(\mat{K}\mat{K}\transp\right)$
also follows the rule defined by~\eqref{eq:avars1}. Furthermore, all diagonal elements of the product~$\mat{K}\mat{K}\transp$ contain the constant value~$n/4$, therefore, the process obtained this way is variance-stationary, unlike the discrete \yperf{} process described in~\cite{Kirkko1}. Proving the above hypotheses rigorously is left for future work, but computer simulations have shown them to hold for at least $R_2,\ldots ,R_{32768}$.

The family of noise sequences presented in this section can provide an alternative view to \yperf{} noise as their Allan variance is exactly constant at certain averaging times and because the sequences are stationary for a given length. The sequences have interesting properties and could be useful in the error propagation analysis of MEMS gyro indexing.

\section{Simulations and Experimental Results}\label{sec:experiments}
In this section, the validity of the calculations presented in Section~\ref{sec:carouseling} is shown by computer simulations and confirmed by experimental results. We first analyze RRW and \yperf{} noise using simulated data, after which the models are applied to authentic data measured by a MEMS gyro.

\subsection{Rate Random Walk Simulation}\label{sec:rrwsimulation}

\begin{figure}
	\centering
	\includegraphics[width=\doublecolumnwidth]{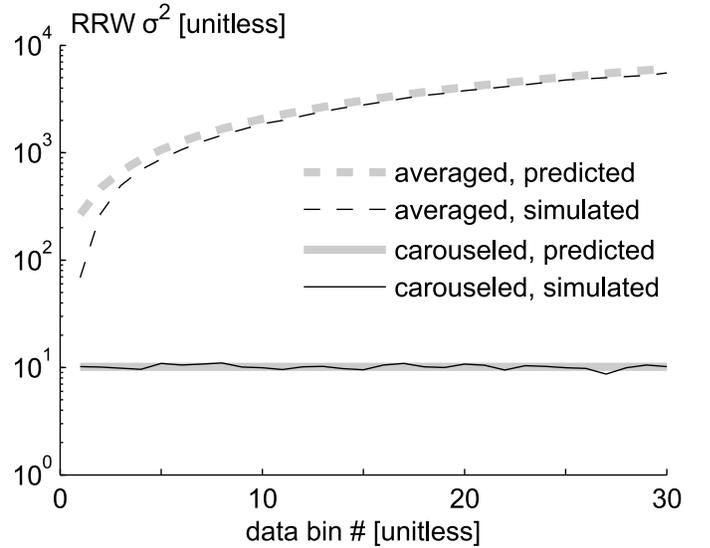}
	\caption{Effect of carouseling on the variance of rate random walk in the computer simulations}
	\label{fig:rrwanalysis}
\end{figure}

The decrease in RRW caused by carouseling was evaluated by first generating $2\times 1000$~mutually independent RRW realizations according to~\eqref{eq:rrw} with driving noise variance~$\varrw = 1$. Then, both the direct averaging and carouseling operations were applied with~$N = 200$, resulting in $1000$~simulation cases. The variances of the resulting sequences are plotted as functions of time (i.e., average block or carouseling revolution number, referred to as data bins) in Fig.~\ref{fig:rrwanalysis} along with the variances predicted by~\eqref{eq:averagecoefficient} and~\eqref{eq:carouselcoefficient}, also including the contribution of the cosine term of~\eqref{eq:consecutivecarousels} in the latter. It can be seen that the predictions match the simulation realizations quite well. The prediction of the averaged variances was computed by neglecting the lower order terms in~\eqref{eq:averagecoefficient}, but this inaccuracy becomes insignificant quickly when the index of the data bin increases.

\subsection{\yperf{} Noise Simulation}\label{sec:yperfsim}

\begin{figure}
	\centering
	\includegraphics[width=\doublecolumnwidth]{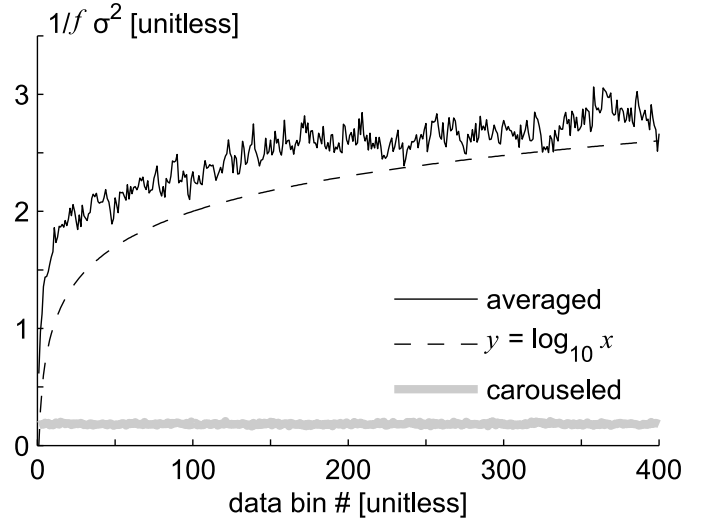}
	\caption{Effect of carouseling on the variance of \yperf{} noise in the computer simulations}
	\label{fig:yperfanalysis}
\end{figure}

A simulation of the evolution of the variance of \yperf{} noise equivalent to that presented in Section~\ref{sec:rrwanalysis} for RRW is shown in Fig.~\ref{fig:yperfanalysis}. The \yperf{} noise sequences were generated according to~\eqref{eq:yperf} with~$\varfl = 1$ and $d = 1/2$, and the averages and carouseling were computed using $N = 200$. Since the $y$-axis scale is linear in this figure, as opposed to Fig.~\ref{fig:rrwanalysis}, it can be seen that the variance of averaged \yperf{} noise increases slowly in comparison with RRW. The rate of increase seems to be logarithmic, which would be natural when considering the relation established in~\eqref{eq:limitcomparisontest}. In contrast, the carouseled \yperf{} noise exhibits no visible increasing trend in Fig.~\ref{fig:yperfanalysis}. Furthermore, its variance is significantly smaller than the driving noise variance~$\varfl = 1$ and a visual inspection shows that the variance is also smaller than that of averaged \yperf{} noise.

\subsection{Real Gyro Data Test}\label{sec:statictest}

\begin{figure}
	\centering
	\includegraphics[width=\doublecolumnwidth]{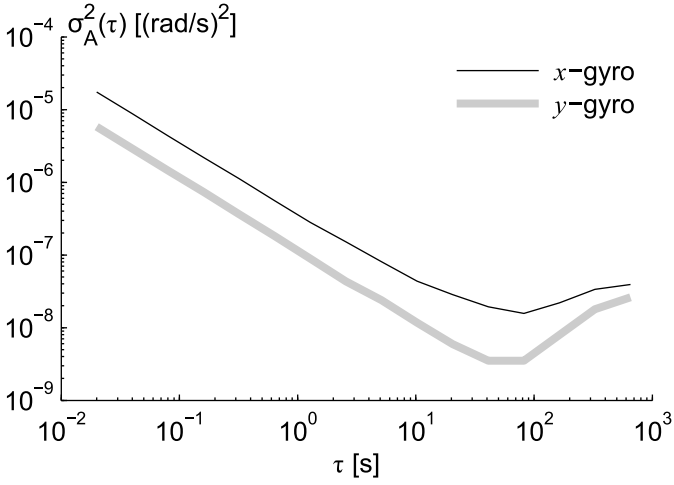}
	\caption{Allan variances of the two gyros used for carouseling}
	\label{fig:avars}
\end{figure}

Test data were recorded for one hour at a sampling rate of~$100$~Hz using a three-axis MEMS gyro~\cite{STgyro}; the Allan variances of the $x$- and $y$-gyro data computed according to~\eqref{eq:avar} are plotted in Fig.~\ref{fig:avars}. During the entire test, the true angular rate to be measured by the sensors was zero. It is interesting to notice that the $x$-gyro exhibited larger variations than the $y$-gyro, but the cause of this discrepancy was not investigated further. The main error parameters for the two gyros are estimated in Table~\ref{tab:gyroerrors}. Since the $x$-gyro does not exhibit a clear ascending RRW slope, its RRW was estimated at the same value of~$\tau$ as for the $y$-gyro. Since the variance of \yperf{} noise was observed to increase very slowly even in the averaged case in Section~\ref{sec:yperfsim}, \yperf{} noise is neglected in this analysis. The bias instability values are only mentioned for reference in Table~\ref{tab:gyroerrors}.

\begin{figure}
	\centering
	\subfloat[]{%
		\includegraphics[width=\doublecolumnwidth]{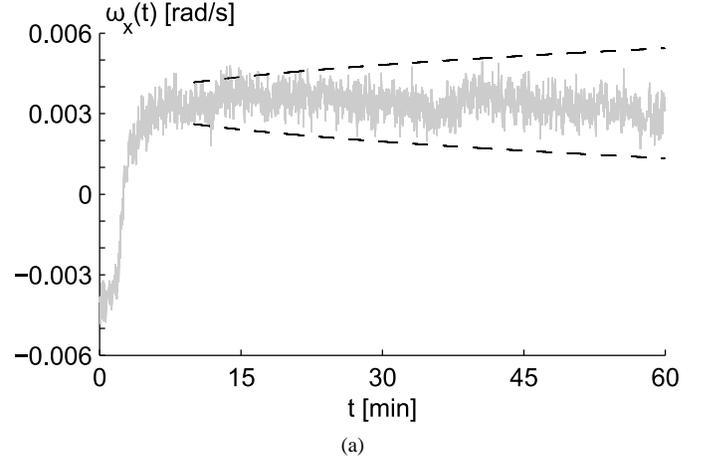}		
		\label{fig:data200:x}
	}\\
	\subfloat[]{%
		\includegraphics[width=\doublecolumnwidth]{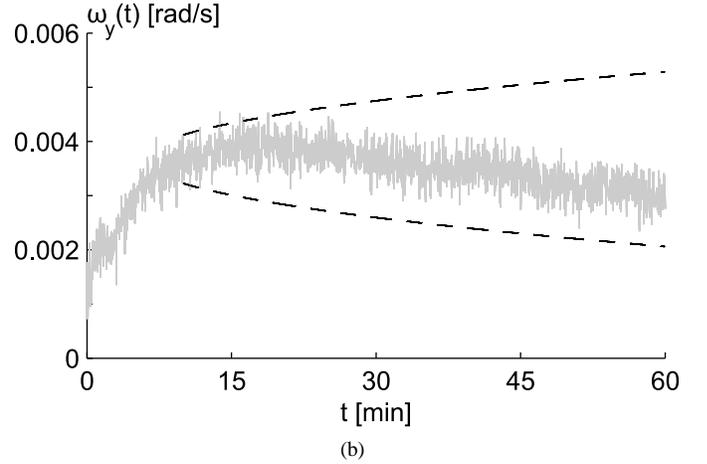}		
		\label{fig:data200:y}
	}\\
	\subfloat[]{%
		\includegraphics[width=\doublecolumnwidth]{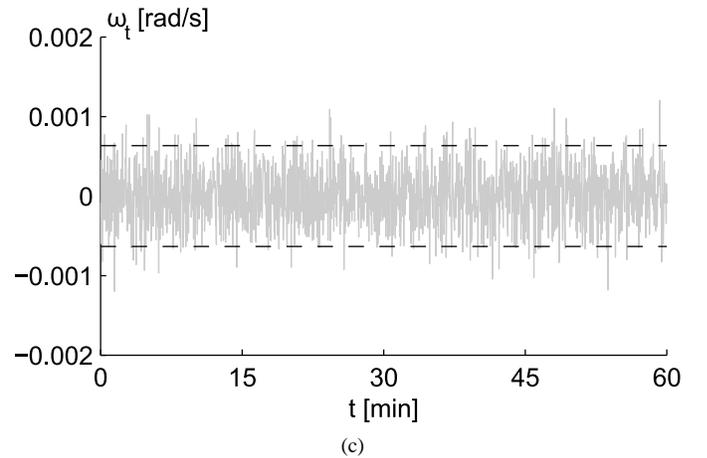}		
		\label{fig:data200:c}
	}
	\caption{Averaged and carouseled gyro data with zero input (solid gray line) and their estimated $2\sigma$~confidence intervals (dashed black line) with $N = 200$: \protect\subref{fig:data200:x}~averaged $x$-gyro; \protect\subref{fig:data200:y}~averaged $y$-gyro; \protect\subref{fig:data200:c}~carouseled estimate}
	\label{fig:data200}
\end{figure}

Fig.~\ref{fig:data200} shows the directly averaged and carouseled data with $N = 200$~samples (consequently, $T = 2$~s) and the respective predicted confidence intervals which are estimated as the sum of white noise and RRW variances, i.e., excluding the contribution of \yperf{} noise. Note that since~\eqref{eq:wnvariance} relies heavily on the assumption of equal white noise variances, the average of the white noise variances of the two gyros was used in the computations. Furthermore, it can be seen that the gyros exhibit a significant initial bias drift, probably due to sensor warm-up. However, this phenomenon is not visible in the carouseled estimate. Errors due to change in ambient temperature or due to aging of the sensor~\cite{Vazuez5469568} are difficult to model, and carouseling clearly makes these errors less effective in the output solution. The initial transient period was excluded from the computation of Allan variance in Fig.~\ref{fig:avars} and the confidence interval estimation in Fig.~\ref{fig:data200} except for the carouseled case. $6.5~\%$~of the carouseled angular rate data points exceed the $2\sigma$ confidence interval which should correspond to $95~\%$ of the samples. Considering that the confidence interval was estimated neglecting the contribution of $1/f$~noise, the result can be regarded as satisfactory.

\begin{table}
	\centering
	\caption{Gyro error parameters as estimated from Fig.~\ref{fig:avars}~\cite{rlgstandard}}
	\label{tab:gyroerrors}
	\begin{tabular}{llrr}
	\toprule
	Parameter & Unit & \multicolumn{2}{c}{Value} \\
	 \cmidrule(lr){3-4}
	  &         &  $x$-gyro & $y$-gyro \\
	\midrule
	$\varwn~(\tau = 1~\mathrm{s})$ & $\left(\mathrm{rad}/\mathrm{s}\right)^2$ & $3\cdot 10^{-7}$ & $1\cdot 10^{-7}$ \\
	$\varrw$ & $\left(\mathrm{rad}/\mathrm{s}\right)^2/\mathrm{s}$ & $3\cdot 10^{-10}$ & $2\cdot 10^{-10}$ \\ 
	Bias instability & $\mathrm{rad}/\mathrm{s}$ & $1\cdot 10^{-5}$  & $6\cdot 10^{-5}$ \\
	\bottomrule
	\end{tabular}
\end{table}

\begin{figure}
	\centering
	\subfloat[]{%
		\includegraphics[width=\doublecolumnwidth]{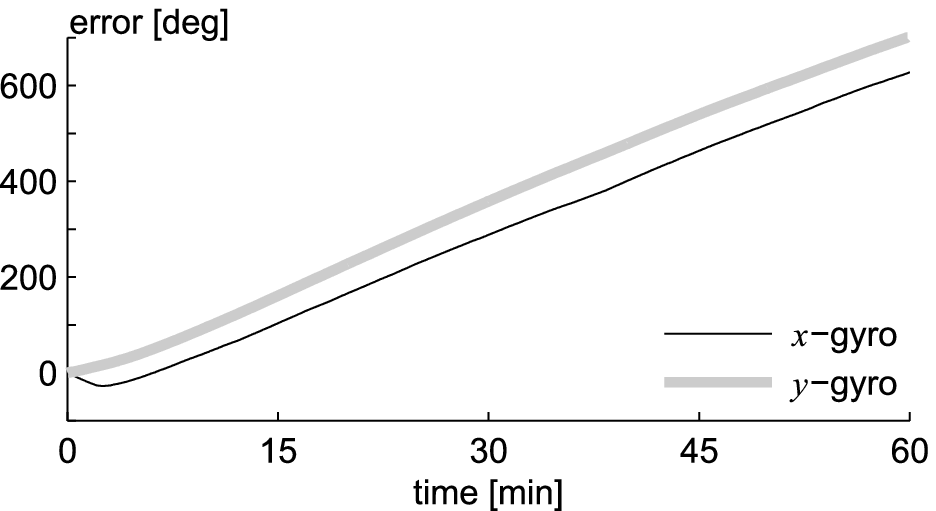}		
		\label{fig:cumsum:avg}
	}\\
	\subfloat[]{%
		\includegraphics[width=\doublecolumnwidth]{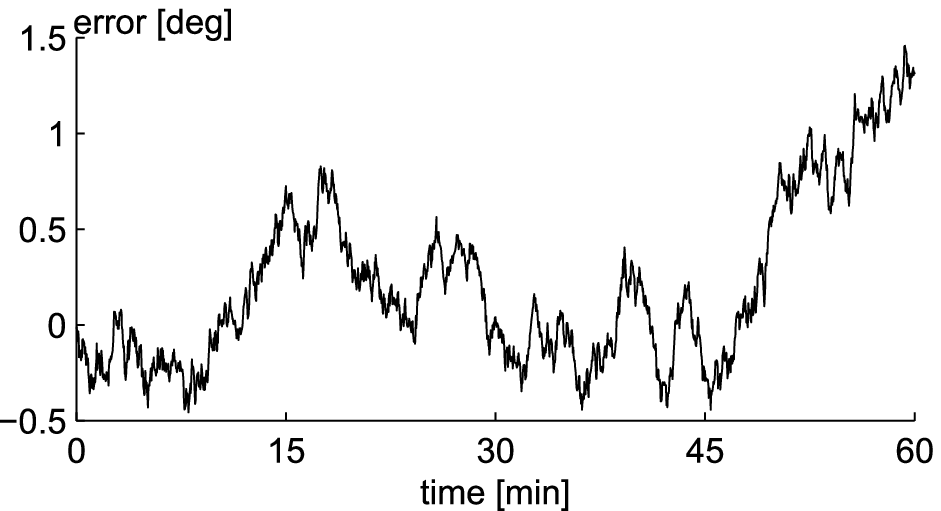}		
		\label{fig:cumsum:car}
	}
	\caption{Accumulated angle estimation errors when integrating the data shown in Fig.~\ref{fig:data200} with $N = 200$, $T = 2$~s: \protect\subref{fig:cumsum:avg}~averaged $x$- and $y$-gyros; \protect\subref{fig:cumsum:car}~carouseled estimate}
	\label{fig:cumsum}
\end{figure}

The effect of carouseling on the resulting angle estimates obtained by integrating the gyro data is illustrated in Fig.~\ref{fig:cumsum}. The increase in accuracy due to carouseling is dramatic: while the directly averaged gyro measurements lead to angle errors exceeding~$600\deg$ after one hour, the carouseled angle errors accumulate to only approximately~$1.5\deg$ in $60$~minutes. The errors shown in Fig.~\ref{fig:cumsum} include the initial warm-up phase where the gyro biases were not yet stabilized.

\subsection{Field Test}
\begin{figure}
	\centering
	\includegraphics[width=.9\doublecolumnwidth]{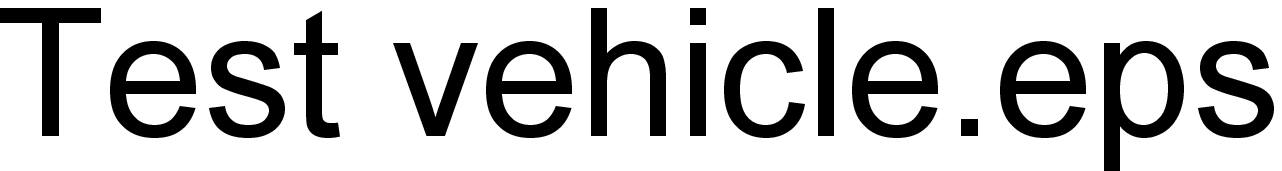}
	\caption{Test vehicle used in field test}
	\label{fig:fig_rearwheel}
\end{figure}

\begin{figure}
	\centering
	\includegraphics[]{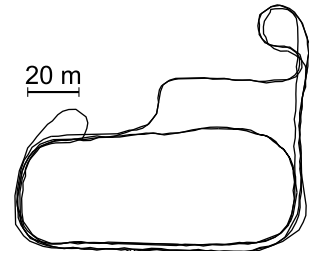}
	\caption{Test route}
	\label{fig:fig_trajectory}
\end{figure}

\begin{figure}
	\centering
	\subfloat[]{%
		\includegraphics[width=.9\doublecolumnwidth]{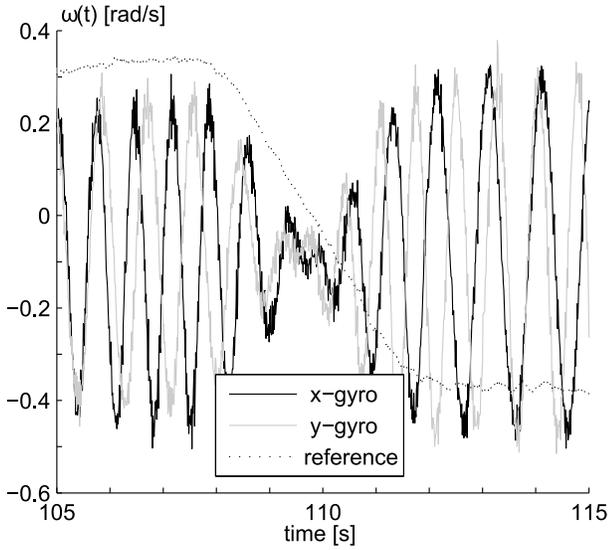}
		\label{fig:raw_wheel}
	}\\%
	\subfloat[]{%
		\includegraphics[width=.9\doublecolumnwidth]{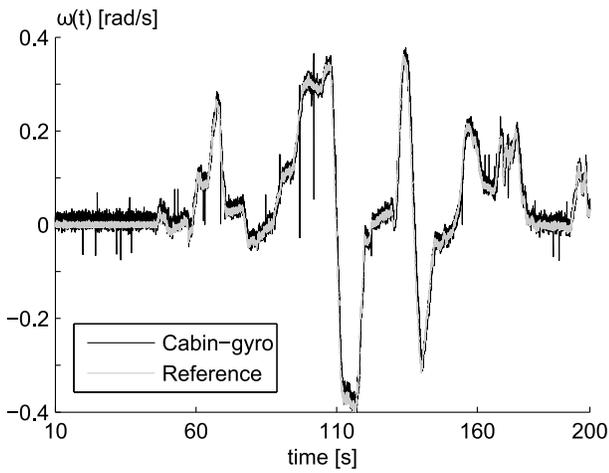}
		\label{fig:raw_cabin}
	}%
	\caption{Raw data: \protect\subref{fig:raw_wheel}~measured at wheel; \protect\subref{fig:raw_cabin}~measured inside the cabin}
	\label{fig:raw_data}
\end{figure}

\begin{figure}
	\centering
	\includegraphics[width=.9\doublecolumnwidth]{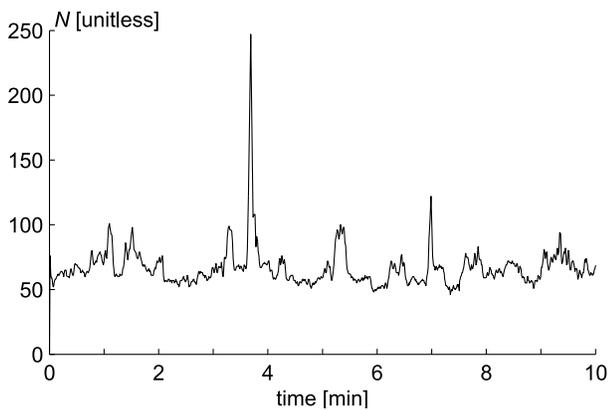}
	\caption{Number of carouseling points~$N$ during the field test}
	\label{fig:N}
\end{figure}

 To show practical application of carouseling we performed a test run with passenger car with an inertial measurement units attached to the right-hand side rear (non-steerable) wheel. The carouseling axis of interest was chosen to be the vertical axis, i.e., heading gyros were considered. The test vehicle, shown in Fig.~\ref{fig:fig_rearwheel}, included the following measurement units:
\begin{itemize}
        \item dual-frequency GPS receiver \cite{Novatel1} and reference ring-laser gyro unit \cite{NovatelBDS} with sub-degree heading accuracy for the test period,
  \item MEMS inertial measurement unit with 3D accelerometer~\cite{STacc} and 3D gyroscope~\cite{STgyro} (identical to the gyro used in the static test in Section~\ref{sec:statictest}) fixed to the right-hand side rear wheel, and
  \item identical MEMS inertial measurement unit fixed to the center console of the vehicle.
\end{itemize}
The carouseling was performed by filtering the wheel-based accelerometer data to estimate $\phi$~\cite{patentti}. The vehicle was driven at a slow speed $10$--$12$~km/h in a parking lot for $12$~minutes. The test route is shown in Fig.~\ref{fig:fig_trajectory} and samples of raw measurements are shown in Fig.~\ref{fig:raw_data}. Since the test area was a public parking lot, a constant velocity was impossible to maintain. Therefore, the number of sampling points~$N$ was not constant during the test but varied from revolution to revolution; the values are plotted in Fig.~\ref{fig:N}. The stationary sections at the beginning and the end of the test are not seen in Fig.~\ref{fig:N}, and test vehicle once had to slow down in order to give way to another vehicle, causing a peak in the value of~$N$.

\begin{figure}
	\centering
	\subfloat[]{%
		\includegraphics[width=.9\doublecolumnwidth]{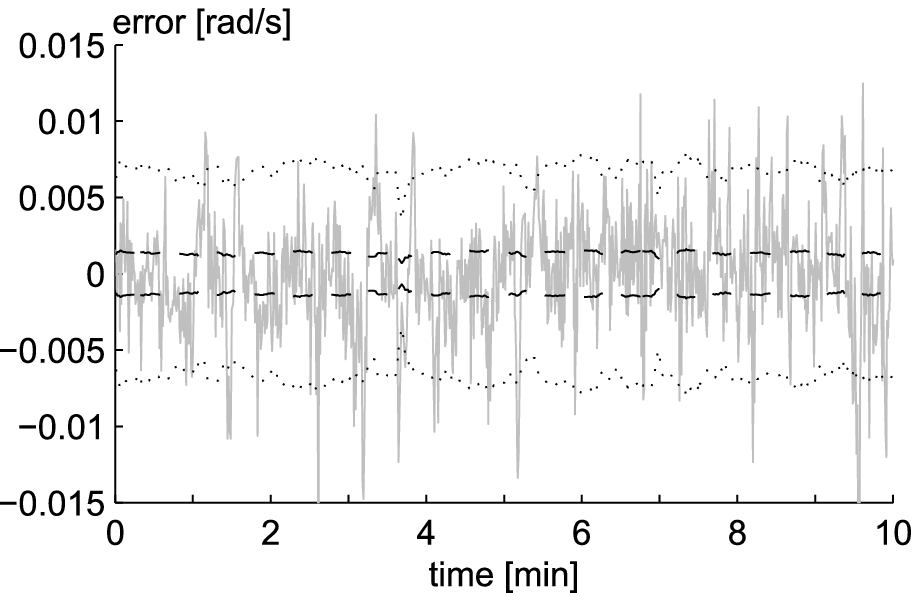}
		\label{fig:fielderror_carousel}
	}\\%
	\subfloat[]{
		\includegraphics[width=.9\doublecolumnwidth]{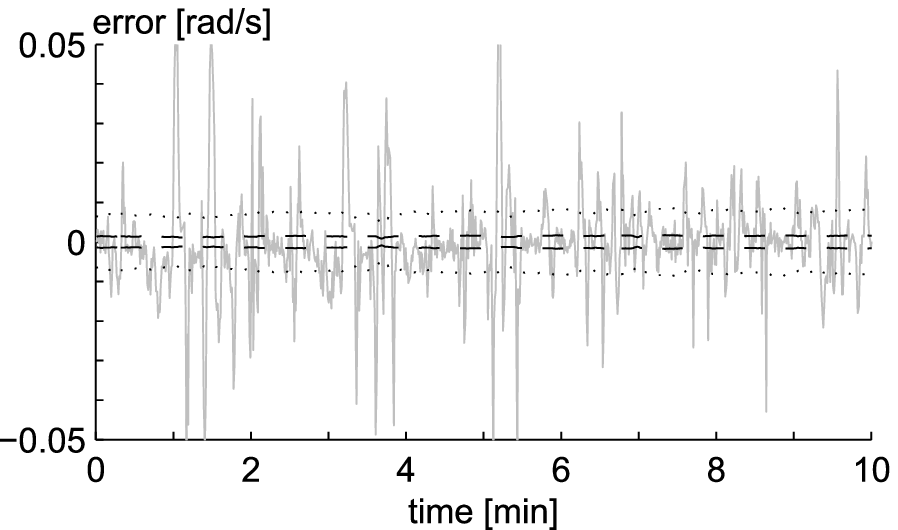}
		\label{fig:fielderror_cabin}
	}%
	\caption{Angular rate estimation error (solid gray line) with $2\sigma$~(dashed) and $10\sigma$~(dotted) intervals predicted based on Table~\protect\ref{tab:gyroerrors}: \protect\subref{fig:fielderror_carousel}~carouseled at wheel; \protect\subref{fig:fielderror_cabin}~averaged inside the cabin}
	\label{fig:fielderror}
\end{figure}

The resulting angular rate estimation errors, as referenced to the ring-laser gyro unit, are plotted in Fig.~\ref{fig:fielderror} along with the predicted confidence intervals. It can be seen that the intervals computed with the error variances corresponding to stationary data do not match the field test errors. There are many possible reasons for the discrepancy. For instance, MEMS sensors are sensitive to accelerations and vibration; the instantaneous carouseling angle~$\phi$ was not precisely known and the carouseling rate was not exactly constant during each revolution; and scale factor and cross-axis sensitivity errors were not compensated for. Nevertheless, the results suggest that the carouseling error equations derived in Section~\ref{sec:carouseling} are correct but the noise variances were not appropriate for the data: there is no drifting trend visible in Fig.~\ref{fig:fielderror_carousel}. Note that the data shown in Fig.~\ref{fig:fielderror} has a lot of local peaks; these are due to a bug in the software that writes the gyro measurements to a memory card.

\begin{figure}
	\centering
	\includegraphics[width=.9\doublecolumnwidth]{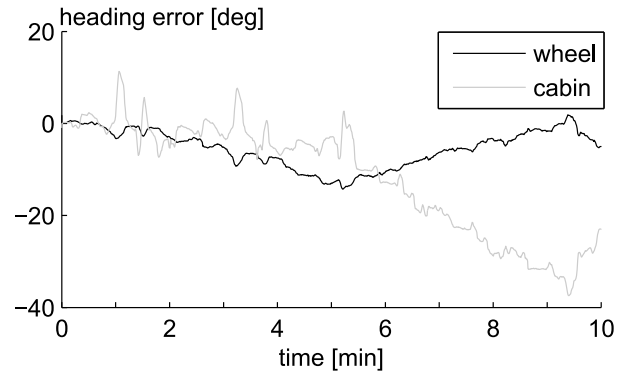}
	\caption{Heading estimation errors with bias compensation for the cabin gyro}
	\label{fig:field_carousel2}
\end{figure}

Errors in the heading estimates obtained by integrating the angular rates are shown in Fig.~\ref{fig:field_carousel2}. A constant additive bias was compensated for in the cabin gyro data; the value of this bias was determined based on the first $30$~seconds of data during which the test vehicle remained stationary. Without such a correction, the cabin gyro heading error increases to $240\deg$ during the test. Although no bias compensation was applied to the wheel-mounted gyro data, the carouseled heading estimates are still more accurate than the cabin-fixed estimates because of the mitigation of rate random walk in the carouseling process. Obviously, the errors encountered in this test are larger than those seen in Section~\ref{sec:statictest} for the reasons discussed above.

\section{Conclusions}\label{sec:conclusions}
In this article, the effect of carouseling on various error processes in the output of a MEMS gyro was studied. It was shown that in addition to  canceling constant biases, carouseling reduces the contributions of rate random walk and \yperf{} noise but does not mitigate white noise better than plain averaging. An immense performance improvement was observed in the case where the gyro outputs are integrated for, e.g., navigation purposes.

As a side product, an alternative approach of synthesizing \yperf{} noise was proposed. The proposed method generates variance-stationary sequences which are not ideal for analyzing the long-time correlation properties of \yperf{} noise, but could be useful, e.g., in the analysis of gyro indexing systems. Investigating the applicability of the noise produced by the method is a topic of future studies.

The variance propagation equations for carouseling were derived under the assumptions of negligible scale factor errors, uniform sensor sampling and carouseling rate, and precise knowledge of the instantaneous carouseling angle. In real-life applications, particularly the last two of these assumptions do not necessarily hold perfectly, as was seen in the field test. Quantifying the sensitivity of the derived covariance prediction formulas to variable slewing rates and multi-axis carouseling, such as the patterns studied in~\cite{Renkoski2008Phd}, is left as future work.


%

%

\ifCLASSOPTIONcaptionsoff
  \newpage
\fi


\bibliographystyle{IEEEtran}
\bibliography{IEEEabrv,main}


\begin{IEEEbiography}[{\includegraphics[width=1in,height=1.25in,clip,keepaspectratio]{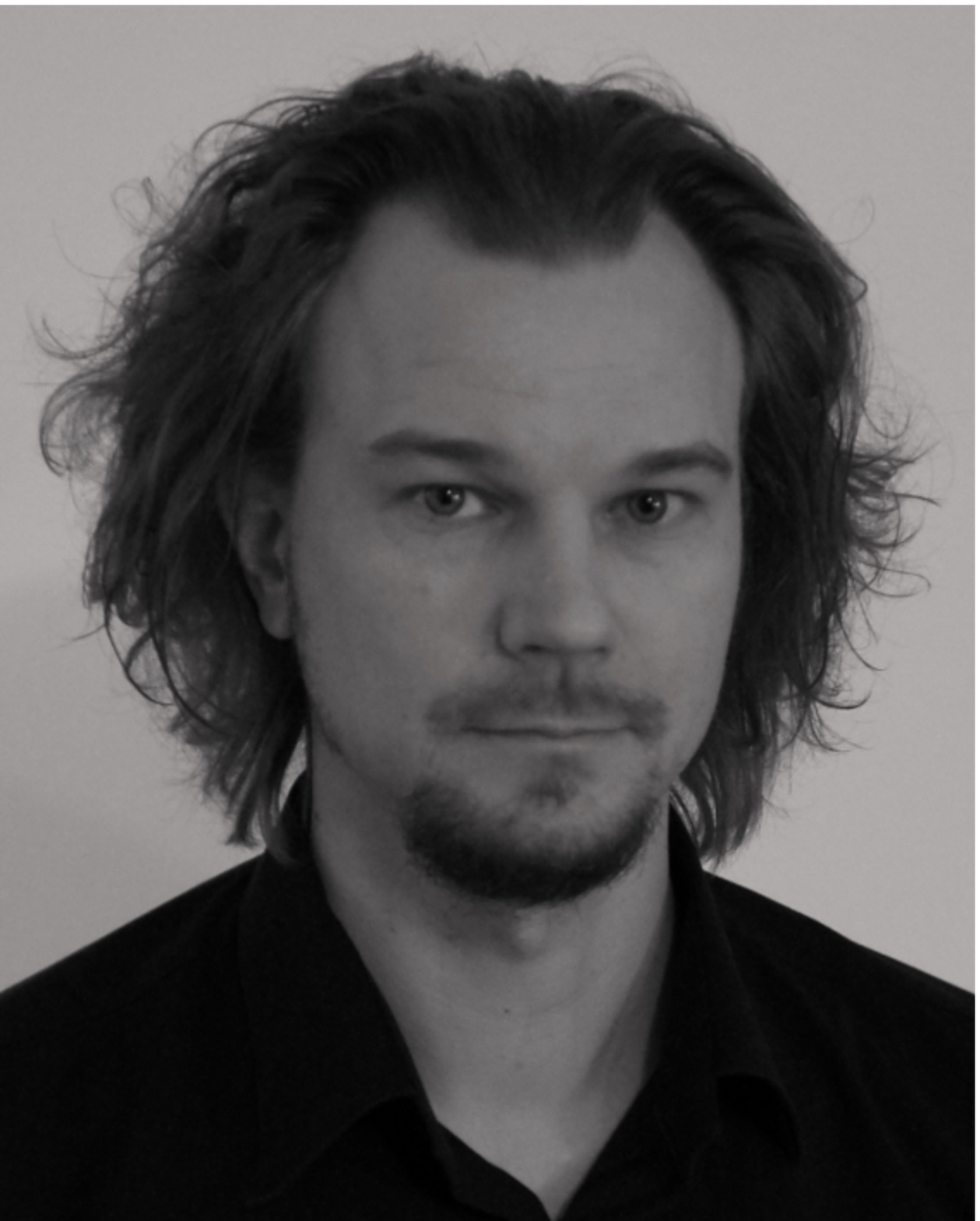}}]{Jussi Collin}
(M'11) received the M.Sc. and Dr.Tech. degrees from the Tampere University of
Technology, Tampere, Finland, in 2001 and 2006, respectively, specializing in
sensor-aided personal navigation. He is currently a Senior Research Fellow with
the Department of Pervasive Computing, Tampere University of Technology. His
research interests include statistical signal processing and novel sensor-based
navigation applications. Dr. Collin is a Vice Chair of the IEEE Finland Section Signal Processing and
Circuits \& Systems Chapter.
\end{IEEEbiography}

\begin{IEEEbiography}[{\includegraphics[width=1in,height=1.25in,clip,keepaspectratio]{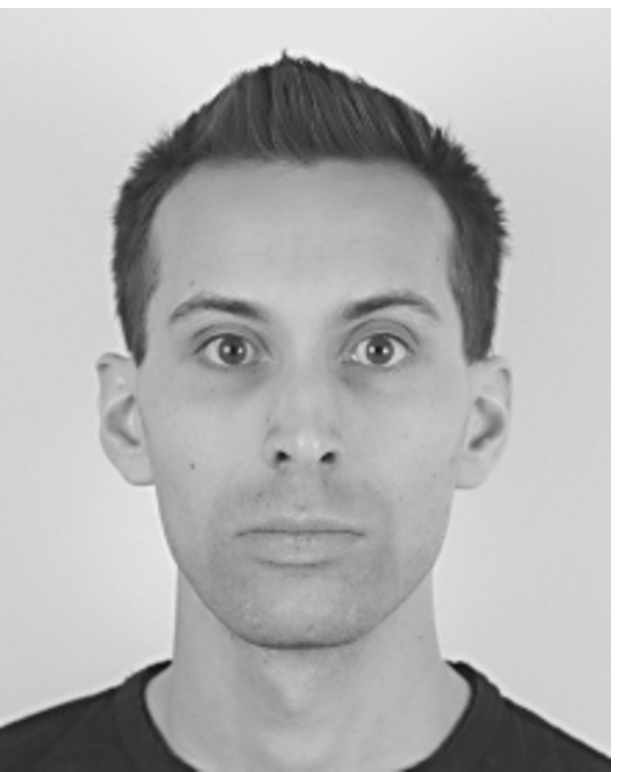}}]{Martti Kirkko-Jaakkola}
(S'12-M'14) \newline received his M.Sc. and D.Sc. (Tech.) degrees from Tampere
University of Technology (TUT), Finland, in 2008 and 2013, respectively. From
2006 to 2013 he worked with the Department of Pervasive Computing, TUT.
Currently, he is a senior research scientist at the Finnish Geodetic Institute,
Kirkkonummi, Finland, where his research interests include precise satellite
positioning, low-cost MEMS sensors, and indoor positioning.
\end{IEEEbiography}

\begin{IEEEbiography}[{\includegraphics[width=1in,height=1.25in,clip,keepaspectratio]{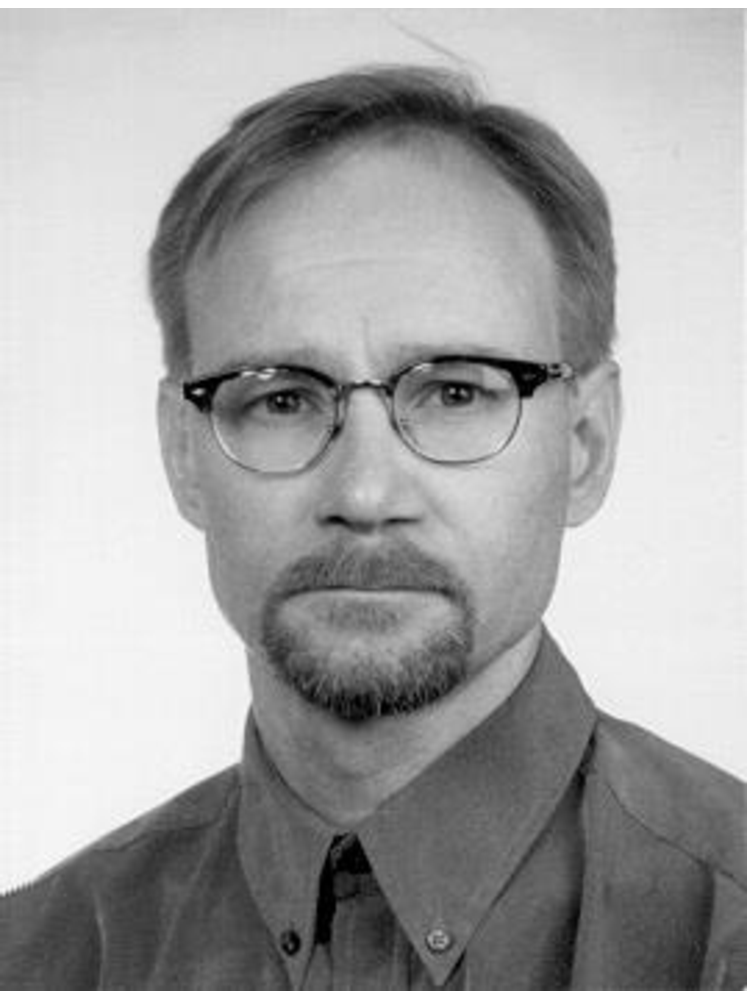}}]{Jarmo Takala}
(S'97­M'99­SM'02) received the M.Sc. (Hon.) degree in electrical engineering and
the Dr.Tech. degree in information technology from the Tampere University of
Technology, Tampere, Finland (TUT) in 1987 and 1999, respectively. He was a
Research Scientist with VTTAutomation, Tampere, from 1992 to 1995. From 1995 to
1996, he was a Senior Research Engineer with Nokia Research Center, Tampere.
From 1996 to 1999, he was a Researcher at TUT. Currently, he is a Professor in
computer engineering at TUT and the Dean of the Faculty of Computing and
Electrical Engineering. His current research interests include circuit
techniques, parallel architectures, and design methodologies for digital signal
processing systems. Prof. Takala is Co-Editor-in-Chief of Journal of Signal
Processing Systems and he was Associate Editor of the IEEE Transactions on
Signal Processing in 2007 - 2011. He was the chair of IEEE Signal Processing
Society's Design and Implementation of Signal Processing Systems technical
Committee in 2011-2013.
\end{IEEEbiography}

\end{document}